\documentclass{aastex}
\usepackage{emulateapj5}

\newcommand{\simgt}{\lower 2pt \hbox{$\, \buildrel {\scriptstyle >}\over {\scriptstyle\sim}\,$}}
\newcommand{\simlt}{\lower 2pt \hbox{$\, \buildrel {\scriptstyle <}\over {\scriptstyle\sim}\,$}}

\newcommand{\pgone}{PG~1115+080}
\newcommand{\apm}{APM~08279+5255}

\newcommand{\chandra}{{\emph{Chandra}}}

\newcommand{\xmm}{\emph{XMM-Newton}}

\slugcomment{Received 2003 March 3; accepted 2003 June 3}
\shorttitle{ Relativistic X-ray BALs from PG~1115+080}
\shortauthors{CHARTAS ET AL.}

\tighten

\begin{document}

\def\sarc{$^{\prime\prime}\!\!.$}
\def\arcsec{$^{\prime\prime}$}
\def\beginrefer{\section*{References}%
\begin{quotation}\mbox{}\par}
\def\refer#1\par{{\setlength{\parindent}{-\leftmargin}\indent#1\par}}
\def\endrefer{\end{quotation}}

\title{ {\sl XMM-NEWTON}  Reveals the  Quasar Outflow in PG~1115+080 }

\author{G. Chartas,\altaffilmark{1} W. N. Brandt,\altaffilmark{1} and S. C. Gallagher\altaffilmark{1,2}}

\altaffiltext{1}{Department of Astronomy \& Astrophysics, Pennsylvania State University,
University Park, PA 16802. chartas@astro.psu.edu, niel@astro.psu.edu, scg@space.mit.edu}

\altaffiltext{2}{MIT Center for Space Research, 77 Massachusetts Avenue, Cambridge, MA 02139.}

\begin{abstract}

We report on an observation of the Broad Absorption
Line (BAL) quasar \pgone\ performed with the \xmm\ observatory.
Spectral analysis reveals 
the second case of a relativistic X-ray absorbing outflow in a BAL quasar.
The first case was revealed in 
a recent observation of \apm\ with the \chandra\ 
X-ray Observatory.
As in the case of \apm, the observed flux of \pgone\ 
is greatly magnified by gravitational lensing.
The relatively high redshift ($z = 1.72$)
of the quasar places the redshifted energies of resonant 
absorption features in a sensitive portion of the \xmm\ spectral response. 
The spectrum indicates the presence of complex low-energy absorption
in the 0.2--0.6~keV observed energy band and high-energy absorption in the 2--5~keV  
observed energy band. The high-energy absorption is best modeled by two Gaussian 
absorption lines with rest-frame energies of 7.4~keV and 9.5~keV. 
Assuming that these two lines are produced by resonant absorption due to
Fe~{\sc xxv} K$\alpha$, we infer that the X-ray absorbers are outflowing with
velocities of $\sim$~$0.10c$ and $\sim$~$0.34c$, respectively. 
We have detected significant variability of the energies and widths
of the X-ray BALs in  \pgone\ and \apm\ over timescales of 19 
and 1.8~weeks (proper-time), respectively. The BAL variability observed from \apm\
supports our earlier conclusion that these absorbers are most likely  
launched at relatively small radii of \simlt\  $10^{16}(M_{BH}/M_{8})^{1/2}$~cm.
A comparison of the ionization properties and column densities of the 
low-energy and high-energy absorbers indicates that these absorbers are likely distinct;
however, higher spectral resolution is needed to confirm this result. 
Finally, we comment on prospects for constraining the kinematic and ionization 
properties of these X-ray BALs with the next generation of X-ray observatories. 

\end{abstract}


\section{INTRODUCTION}

Most of our current understanding of the kinematic structure  in
Broad Absorption Line quasar (BALQSO) outflows stems from 
numerous studies of the velocity structures 
of BAL features that appear blueward of resonant UV emission lines.
The similarities between the emission-line properties of
quasars with and without BAL features 
led Weymann et al. (1991) to conclude that BALQSOs are ``normal'' quasars 
with a BAL global covering factor of $\sim$ 10\%,
viewed along lines of sight that traverse the outflowing UV absorbers.
In this model, quasar winds exist in most quasars; however, because of the 
relatively small opening angles of these outflows only a small fraction of radio-quiet quasars 
have detectable BAL features in their UV and/or optical spectra. 
Spectropolarimetric observations, the statistics of gravitationally lensed BALQSOs, 
improved assessments of incompleteness, and
theoretical arguments suggest that present optical surveys may be biased against
the detection of BALQSOs (e.g., Goodrich 1997; Chartas 2000; Krolik \& Voit 1998;
Tolea, Krolik, \& Tsvetanov \ 2002; Hewett \& Foltz 2003).
These analyses indicate that the true fraction of BALQSOs and BAL 
covering factors may be substantially larger (\simgt\  20\%) than the often quoted value 
of $\sim$ 10\%. 

The study of BALQSOs in the X-ray band may provide valuable information
regarding the total column densities and the ionization equilibrium of the BAL absorber.
X-ray absorption in BALQSOs is expected to be largely due to 
photoelectric, bound-free absorption
and is less sensitive to effects such as saturation and the degree of ionization,
often encountered in optical and UV observations (e.g., Arav 1998).
Because of the heavy X-ray absorption in BALQSOs, 
the X-ray spectra obtained to date have modest (e.g., PG 2112+059) to poor 
signal-to-noise (S/N) ratios and cannot accurately constrain the column densities,
ionization state, geometry, or kinematic properties of BAL absorbers 
(e.g., Gallagher et al. 2001; Green et al. 2001; Gallagher et al. 2002). Furthermore, 
it is not clear whether the X-ray and UV absorbers are distinct.

Our recent \chandra\ observation of the BALQSO \apm\ indicated that 
quasars may contain winds with mass outflow rates considerably larger 
than previously thought (Chartas et al. 2002). This was 
suggested by the identification of two X-ray absorption features at 
rest-frame energies of 8.1 and 9.8~keV. By interpreting these features as resonant
absorption lines of highly ionized Fe (e.g., Fe~{\sc xxv} K$\alpha$),
we inferred the presence of X-ray BAL material outflowing from the 
central source at velocities of $\sim$ 0.2$c$ and $\sim$ 0.4$c$. 
Our analysis of \apm\ also suggested that the X-ray BAL material in this object is launched 
at radii of $ \simlt\  2 \times 10^{17}$~cm and therefore is likely not 
associated with the UV BAL material which is thought to be launched at larger radii. 
An observation of \apm\ performed with \xmm\ $\sim$ 1.8~weeks (proper-time) 
after the \chandra\ observation showed an absorption feature near 7.4~keV (rest-frame) that was interpreted by 
Hasinger, Schartel, \& Komossa (2002, hereafter H02) as a highly ionized 
iron edge  (Fe~{\sc xv}--Fe~{\sc xviii}). 
If this interpretation is accepted an overabundance of Fe  
(factor $\sim$ 2--5) is required to fit the \xmm\ spectrum, assuming that the 
observed low and high-energy
absorbers are the same.
Recently, an ionized outflow with a velocity of 
0.08--0.1$c$ has also been detected in \xmm\
observations of the narrow emission-line quasar PG~1211+143 (Pounds et al. 2003).
Interestingly, there are no detected UV BALs from PG~1211+143. 
Additional insight into the association between the X-ray and UV BAL material
was recently provided by  contemporaneous \chandra\ and $\it HST$ STIS observations 
of the BALQSO PG~2112+059 (Gallagher et al. 2003).
Significant differences in the variability of the X-ray and UV absorbers in PG~2112+059 
suggest the distinct nature of the X-ray and UV BAL material.

The confirmation of relativistic X-ray absorbing outflows in most quasars 
would imply higher mass-outflow rates than those based on
outflow velocities derived from UV BALs.
Such outflowing winds could provide 
a significant contribution to the enrichment of the 
vicinities of the quasars and their host galaxies with high-metallicity material. 
To address the question of whether relativistic X-ray absorbing
outflows are a common phenomenon in quasars, 
we performed a deep observation of \pgone\ ($z = 1.72$, {\it B} = 16.1) with \xmm. 
\pgone\ is a well-suited object for probing the kinematics of quasar winds. 
The relatively small velocity spread observed 
for the UV absorption in \pgone,
compared to the typical range of 5,000 -- 25,000~km s$^{-1}$
observed in BAL quasars, suggests that \pgone\ be classified as
a mini-BALQSO (Turnshek 1988; Barlow, Hamann, \& Sargent 1997).

The large flux magnification, estimated to be a factor of $\sim$ 20--40 
(e.g., Impey et al. 1998), produced by gravitational lensing 
provided a bright background source that allowed the detection of
absorption features from the outflowing wind.
Here we present constraints on the kinematic and ionization properties of the outflowing wind of \pgone\
and provide results from our investigation of variability of the X-ray BALs
of  \pgone\ and \apm. Throughout this paper we adopt a 
$\Lambda$-dominated Universe with $H_{0}$ = 70~km~s$^{-1}$~Mpc$^{-1}$, 
$\Omega_{\Lambda}$ = 0.7, and  $\Omega_{M}$ = 0.3.

\section{OBSERVATIONS AND DATA ANALYSIS}

\pgone\ was observed for 62.6~ks on 2001 Nov 25 with \xmm. It was also 
observed with \chandra\ on 2000 June 2 and 2000 Nov 3 for 26.8~ks and 10.0~ks, respectively. 
The spectral analysis of the \chandra\ observations were presented in
Gallagher et al. (2002).
In \S 3.2, we provide a comparison between the \xmm\ and \chandra\
spectral results and include an update on the \chandra\ analysis of \pgone\  that  
incorporates a recently developed contamination model for the ACIS instrument. 

No significant background flares were present during the \xmm\ observation,
and the average background rates for the 
European Photon Imaging Camera (EPIC) PN and MOS detectors in the 0.2--10~keV band  
were 4.4 $\times$ 10$^{-6}$ events~s$^{-1}$~arcsec$^{-2}$
and  8.3 $\times$ 10$^{-7}$ events~s$^{-1}$~arcsec$^{-2}$, respectively.
We analyzed the data with the standard analysis software SAS 5.3.  
\pgone\ was clearly detected with the PN,
MOS1, and MOS2. The total number of source counts detected in the 
0.2--10~keV band was $\sim$ 12,200~counts  for the PN detector
and $\sim$ 8,185~counts for the combined 
MOS1 and MOS2 detectors (henceforth referred to as MOS1+2).
Flux variability was detected at the $ > $ 90\%  confidence level
based on a Kolmogorov-Smirnov (K-S) test. 
The timescale of the flux variations is of order 1000~s.
A description of the timing analysis and inferred constraints on
the Hubble constant based on the estimated time delay between the 
lensed images is presented in a separate paper (Chartas et al. 2003).
We filtered the PN and MOS data by selecting events with \verb+PATTERNS+
in the 0--4 and 0--12 ranges, respectively.
The MOS1 and MOS2 event lists were combined using the SAS task \verb+merge+
to increase the S/N ratio of the extracted spectrum.

The extracted spectra of \pgone\ from the PN and MOS 
were grouped to obtain a minimum of 100 counts in each energy bin,
allowing use of $ \chi^{2}$ statistics.
Background spectra for the PN and MOS1+2 detectors were extracted
from source-free regions near \pgone.
The PN and MOS1+2 spectra were then fitted 
simultaneously with a variety of models employing 
\verb+XSPEC+ version 11.2 (Arnaud 1996).  
We tested the sensitivity of our results to the selected background and source extraction
regions by varying the locations of the background regions and varying the 
sizes of the source extraction regions. We did not find any significant change in the 
background-subtracted spectra.  For all models we 
included Galactic absorption due to neutral gas with a 
column density of $N_{\rm H}$=3.53 $\times$ 10$^{20}$~cm$^{-2}$ (Stark et al. 1992).
All quoted errors are at the 90\% confidence level unless mentioned otherwise
with all parameters taken to be of interest except absolute normalization.

We first simultaneously fitted the PN and MOS1+2  
spectra of \pgone\ with a model consisting of a power law with neutral intrinsic
absorption at $z = 1.72$ (fit 1 of Table 1). 
This fit supports the presence of an intrinsic absorber with
a column density of  $N_{\rm H}$= 2.7$_{-0.5}^{+0.5}$ $\times$ 10$^{21}$~cm$^{-2}$.  
The fit is not acceptable in a statistical sense [reduced $\chi^{2}$ = 1.54 for 209
degrees of freedom (dof)]. The fit residuals show significant absorption 
from 2--5~keV that contributes to the unacceptable fit.
To illustrate the presence of these features and the absorption features
below 0.6~keV, we fit the spectrum between 1--2.6~keV and 
5.2--10~keV with a power-law model
modified by Galactic absorption and extrapolated this model to the energy ranges not fit
(see Figure 1a).
For clarity we only show the higher S/N ratio PN data in Figure~1; however, all fits
were performed simultaneously to the PN and MOS1+2 data.
In Figures 2a and 2b we show the $\Delta\chi$ residuals
between the best-fit absorbed power-law model
and the PN and MOS1+2 spectra. 
The complex absorption features from 2--5~keV are detected
in both the PN and MOS1+2 spectra. 

We proceeded to model the 2--5~keV features with a variety of models. 
We first considered an absorbed power-law model with two 
cosmologically redshifted Gaussian absorption lines
near the absorption features appearing at energies of ~2.7 and 3.6~keV (observed-frame).
This fit is a significant improvement compared to the previous one at the $ > $ 
99.99\% confidence level and yields a reduced $\chi^{2}$ = 1.33 for 201 dof (fit 2 of Table 1).  
The addition of the two Gaussian lines significantly 
reduced the residuals, and the remaining most significant contribution
to the large value of  $\chi^{2}$ arises from the residuals below 0.6~keV. 
These low-energy residuals are commonly detected in 
moderate S/N ratio X-ray spectra of BALQSOs 
and are thought to arise from ionized or partially covering intrinsic 
absorption (e.g., Gallagher et al. 2002).
To test this we replaced the neutral absorber in our spectral model
with an ionized intrinsic absorber (see fit 5 of Table 1). 
In particular, we used the \verb+absori+ model contained in 
\verb+XSPEC+ (Done et al. 1992). In Figure~1b we show
the PN spectrum fit with this model and the residuals of this fit.
We found a significant improvement in fit quality at the $ > $ 99.99\% 
confidence level (according to the $F$-test)
with reduced $\chi^{2}$ = 1.16 for 200 dof (see fit 5 of Table 1). 
The best-fit rest-frame energies and widths  
of the absorption features are $E_{abs1} = 7.38_{-0.10}^{+0.17}$~keV,  
$\sigma_{abs1}$ $ < $ 0.44~keV and $E_{abs2} = 9.5_{-1.4}^{+0.9}$~keV (68\% confidence),
$\sigma_{abs2} = 2.4_{-0.9}^{+1.2}$~keV (68\% confidence).
\footnote{The unusually large width of the
absorption line at 9.5~keV may be due
to multiple unresolved narrower components. 
Higher resolution spectra are needed to resolve this issue.}

Both of the 2--5~keV features occur in a well-calibrated energy region 
where the effective area
of the combined telescope and instrument responses varies smoothly. The best-fit 
value for the photon index is $\Gamma$ = 1.90$_{-0.04}^{+0.04}$,
consistent with those of typical radio-quiet quasars.
The ionized absorber is characterized by an ionization parameter of
$\xi = L/nr^{2} = 101.4_{-48}^{+58}$~erg~cm~s$^{-1}$ and a hydrogen column density
of $N_{\rm H} = 1.1_{-0.4}^{+0.5}$ $\times$ 10$^{22}$~cm$^{-2}$, 
where $L$ is the integrated 5~eV--300~keV incident luminosity, $n$ is the
number density of the absorber, and $r$ is the distance between the 
absorber and ionizing source.

We also modeled the absorption from 2--5~keV
with one and two absorption edges. 
The fit with one edge is slightly worse (at the 90\% confidence
level) than the fit that included the two Gaussian lines and yields 
a reduced $\chi^{2}$ = 1.18  for 206 dof (fit 3 of Table 1).
The best-fit edge energy and optical depth are $E_{edge}$ = 7.07$_{-0.23}^{+0.20}$~keV
and $\tau_{edge}$ = 0.31 $\pm$ 0.12.  
An inspection of this fit  shows that significant residuals are
present from 2--5~keV.
The fit with two absorption edges is comparable in quality to the
fit that included two Gaussian absorption lines and yields a 
reduced $\chi^{2}$ = 1.16 for 204 dof.
The best-fit energies and optical depths of the two absorption edges are 
$E_{edge1}$ = 7.03$_{-0.32}^{+0.19}$~keV,  $\tau_{edge1}$ = 0.31 $\pm$ 0.11
and $E_{edge2}$ = 9.63$_{-0.43}^{+0.77}$~keV, $\tau_{edge2}$ = 0.22 $\pm$ 0.14.
The best-fit values for the edge energies would imply 
that the first edge arises from nearly neutral Fe with a $2\sigma$ confidence range from
Fe~{\sc I} to Fe~{\sc vii}, whereas the second edge energy 
lies $1\sigma$ above the threshold energy of the highest ionization level of Fe~{\sc xxvi} 
(9.278~keV). While the fit that incorporates two absorption edges is acceptable 
in a statistical sense, the inferred energies and optical depths 
are difficult to explain with a physically acceptable model (see \S 3.1).
The multiple-edge model would imply that the detected high-energy absorption is produced 
by two distinct absorbers, one nearly neutral and the other highly ionized.
The optical depths of the edges imply hydrogen column densities 
of $\sim$ 2.7 $\times$ 10$^{23}$ cm$^{-2}$ for
the edge at 7.03~keV and $\sim$ 6.9 $ \times$ 10$^{23}$ cm$^{-2}$
for the edge at 9.63~keV (for solar abundances and adopting the threshold cross 
sections of Verner \& Yakovlev 1995).
In \S 3.1 we investigate whether spectral models that incorporate 
direct and scattered radiation from the central source can provide 
acceptable fits to the low and high-energy absorption features in the \xmm\ spectrum of
\pgone.

\section{DISCUSSION}
Our \xmm\ observation of \pgone\ has revealed a BALQSO with
two absorption features very similar in rest-frame energy to the ones recently 
identified in \chandra\ observations of the BALQSO \apm. 
In \S 3.1 we investigate three spectral models
to explain these high-energy absorption
features, provide constraints on their kinematic and ionization properties,
and examine a possible connection between the low and high-energy 
absorbers.  In \S 3.2 we investigate possible variability of these 
absorption features by comparing  the \chandra\ and \xmm\ observations of
\pgone\ separated by 19~weeks (proper-time). We also briefly discuss 
the variability of similar absorption features 
detected in \chandra\ and \xmm\ observations of \apm\ separated by 1.8~weeks (proper-time) 
and place constraints on the size and location of these absorbers.

\subsection{Origin and Properties of  the X-ray Absorbers in \pgone}

We attempted to determine the origin of the low and high-energy absorption features 
seen from \pgone\  
by considering the three following models:


(a) Our first model consisted of neutral and ionized absorption with 
variable metal abundances. 
The neutral component consisted of a power law modified by the
\verb+XSPEC+ model \verb+zvphabs+ which assumes photoelectric
absorption with variable abundances.
The ionized component consisted of a power law modified by the \verb+XSPEC+ model
\verb+absori+. The Fe abundances of both components were allowed to vary 
during the spectral fits. 
This multi-component model was an attempt to 
account for the results of the spectral fits that assumed a two-edge model 
(described in \S 2) to fit the high-energy absorption features.
We assumed that our lines of sight through the neutral and ionized absorber are different. 
This can occur if the observed spectrum consists
of (1) direct absorbed and (2) scattered, absorbed emission from the central source.
The fit with this multiple-component model was acceptable in a statistical sense
yielding a reduced $\chi^{2}$ = 1.13 for 202 dof.
However, there are problems with 
explaining the resulting values for several of the parameters.
First, the best-fit value of the Fe abundance needed to model 
the neutral absorption feature near $\sim$ 7.1~keV 
is $\sim$ 425$_{-330}^{+335}$ times solar, and the Fe abundance needed to model
the ionized low-energy absorption is $\sim$ 15 $\pm$ 10 times solar 
(errors are at the 90\% confidence level). 
These values are unphysically large.
Second, the ${\Delta}{\chi}$ residuals of this fit indicate that
the double line of sight    
model can fit the low-energy absorption well,
but, significant residuals remain near the 
absorption feature at 9.6~keV. 
We considered the case where the ionized and neutral absorbers
were along the same line of sight.
The fit with this model was not acceptable in a statistical sense with a
reduced $\chi^{2}$ = 1.26 for 205 dof.
The best-fit value of the column density of the neutral
absorber for this model is $N_{\rm H}$ = 0.0$_{-0.0}^{+0.1}$~$\times$~10$^{21}$~cm$^{-2}$.

(b) We also investigated whether intervening absorbers in the lens galaxy at $z = 0.31$
or in possible damped Lyman alpha systems
along the line of sight could explain the high-energy absorption features.
Spectroscopic observations of \pgone\ in the UV
have not detected any Lyman-limit discontinuities
(Green et al. 1980; Tripp et al. 1990)
and place an upper limit on the lens galaxy's column density of neutral hydrogen of
$N_{\rm H}$ \simlt 3 $\times$ 10$^{16}$ cm$^{-2}$.
The most abundant elements with K-shell threshold energies near
the energies of the observed high-energy absorption features
are Si and S. For the purpose of this investigation 
we compared the K-shell threshold energies of Si and S  (corrected
for the redshift of any intervening absorber between us and \pgone) to
the best-fit energies of $E_{edge1}$ = 2.59~keV
and $E_{edge2}$ = 3.54~keV (observed-frame) obtained from our fits that assumed two
absorption edges (see \S 2 and fit 4 of Table 1).
The energies of the Si {\sc i } -- {\sc xiv} and S {\sc i} -- {\sc xvi}
K-shell threshold energies
range between 1.846--2.673~keV and 2.477--3.494~keV,
respectively. We expect most of the Si and S in intervening
damped Lyman alpha systems or in the lens galaxy
to be neutral or mildly ionized. 
We find that all the K-shell threshold energies of Si and S
fall significantly below the energies
of the observed high-energy absorption features
for absorption produced in the lens galaxy
with the exception of S {\sc xvi}
which falls between the two observed high-energy absorption features.
All ionization levels of Si and S are also
ruled out as producing the absorption edge
at 3.54~keV for any intervening absorber.
We conclude that the observed high-energy absorption features
cannot be produced by absorption in intervening
systems or in the lens galaxy, and the most likely
origin is intrinsic absorption by highly ionized iron.

(c) Finally, we examined a model that assumes the high-energy absorption features
arise from different zones of a highly ionized outflowing wind.
A similar model was used to explain the high-energy absorption
features in the \chandra\ spectrum of \apm.  In particular, for the case of \apm\ 
we interpreted these features as two distinct absorption systems in the quasar outflow with velocities of 
$\sim$ $0.2c$ and $\sim$ $0.4c$, respectively; however, 
see H02 for a different interpretation of the \xmm\ spectrum of \apm. 
For the \xmm\ observations of \pgone\ we found that the absorption features are located 
at rest-frame energies of $7.38_{-0.10}^{+0.17}$~keV (90\% confidence)
and $9.50_{-1.40}^{+0.90}$~keV (68\% confidence); see fit 5 of Table 1.
Applying similar arguments to infer the nature of the lines in \pgone\ as those 
presented in Chartas et al. (2002) for \apm, we conclude that the absorption lines
detected from \pgone\ are most likely due
to highly ionized Fe ions in the quasar wind. 
We estimate that the 7.38~keV and 9.50~keV
absorption features correspond to wind velocities (depending on the ionization state of Fe)
of $0.10c$~(Fe~{\sc xxv}~K$\alpha$),
$0.04c$~(Fe~{\sc xxvi}~K$\alpha$) and $0.34c$~(Fe~{\sc xxv}~K$\alpha$),
$0.30c$~(Fe~{\sc xxvi}~K$\alpha$)
respectively, relative to the systemic redshift of $z = 1.72$.
For these velocity calculations we included the special relativistic
velocity correction.

At the 1$\sigma$ level, the ratio of the two absorption-line 
energies $(E_{abs1}/E_{abs2})$ is inconsistent with the ratio of 
the energies of the iron K$\alpha$ to K$\beta$ transitions of Fe~{\sc xxv}  and Fe~{\sc xxvi}.  
The predicted ratio of $EW_{abs2}/EW_{abs1}$ of the two high-energy absorption lines
is $\sim$ 7.1 (7.4) assuming that these lines are the iron K$\alpha$
and K$\beta$ transitions of Fe {\sc xxv} (Fe {\sc xxvi}) and the absorption
is on the linear part of the curve of growth. 
The observed ratio of $EW_{abs1}/EW_{abs2}$ = 0.1 $\pm$ 0.06
is much less than the predicted ratio, and thus does not obviously support the iron K$\alpha$
and K$\beta$ interpretation of the high-energy absorption lines.
We note, however, that in the likely case where the absorption lines are saturated 
or are composed of multiple unresolved components 
the observed $EW_{abs2}/EW_{abs1}$ ratio could deviate significantly from
the predicted ratio assuming the iron K$\alpha$ and K$\beta$ interpretation.
A second argument against the iron K$\alpha$
and K$\beta$ interpretation is the observed radical difference of the line profiles of the absorption
lines at 7.38~keV and 9.50~keV (see fit 5 of Table 1) that is not expected if
these absorption features are produced by iron K$\alpha$ 
to K$\beta$ transitions of Fe~{\sc xxv} or Fe~{\sc xxvi}. 

UV spectra of \pgone\ acquired with the International Ultraviolet Explorer $(IUE)$
reveal an O~{\sc vi}  $\lambda$1033 resonant emission line
accompanied by absorption in the line core and absorption features  blueward of the line that 
extend over a velocity range of 0--6000~km~s$^{-1}$ (Michalitsianos \& Oliverson 1996).
A possible explanation for the large differences between the UV and X-ray BAL wind velocities 
is that the two absorbers are launched at different distances from the central ionizing source 
and are therefore accelerated to different terminal velocities.
To examine this hypothesis further we estimated the wind velocity as a function of radius
following \S 5 of Chartas et al. (2002).
We assumed a minimum launching radius for the UV BAL absorber
of $r_{\rm min} \sim 1 \times 10^{16}(M_{bh}/M_{8})^{1/2}$~cm,
based on hydrodynamical simulations performed by Proga et al. (2000),
where $M_{bh}$ is the mass of the black hole of \pgone\ and $M_{8} = 10^{8} M_{\odot}$.
In Figure 3 we plot the velocity of the UV absorber versus radius from the central 
source for launching radii of $r_{\rm min},  2.5r_{\rm min}$ and $5r_{\rm min}$.
We have assumed a force multiplier of ${\Gamma}_{f}$ = 100 (e.g., Arav, Li, \& Begelman 1994), 
$L_{UV}$ = 5 $\times$ 10$^{44}$ erg s$^{-1}$,
$L_{Bol}$ = 2 $\times$ 10$^{45}$ erg s$^{-1}$, and $L_{Bol}$/$L_{Edd}$ = 0.1.
The terminal velocity of UV material launched at the minimum
launching radius is $\sim$ 0.07$c$, significantly less than the velocities of the X-ray
BAL absorbers. As we proposed in the case of \apm, a possible explanation for the 
larger velocities of the X-ray BAL absorbers is that this material is initially 
accelerated at smaller 
launching radii than the UV BAL material. The short recombination timescales of 
Fe~{\sc xxv} and/or Fe~{\sc xxvi} compared to the estimated time for the X-ray BAL 
material to be accelerated 
close to its terminal velocity imply that the observed highly ionized 
Fe is located near the launching radius where it
would still be exposed to intense ionizing radiation from the central X-ray
source. One of the implications of small launching radii for the X-ray BAL material 
is possible variability over timescales of weeks to months 
of the energies and widths of the X-ray BALs 
as the ionized X-ray absorbers are accelerated to their terminal velocities.

We investigated a possible connection between the low and high-energy X-ray absorbers
by comparing their kinematic, ionization, and absorbing properties. 
We searched for a possible outflow velocity of the low-energy absorber by
allowing the redshift parameter of the ionized-absorber model to be free. 
The best-fit redshift of the absorber 
of $z$ = 1.53$_{-0.7}^{+0.6}$ (90\% confidence errors) was consistent 
with the systemic value of $z$ = 1.72 (and also with the blue-shifted values 
of $z$ = 1.46 and 0.92 expected from the velocities of the high-energy absorbers),
and we found no significant improvement in the low-energy fit residuals and
no improvement in the global fit. 
We note, however, that if the low-energy absorber has a large velocity dispersion
it would be difficult to detect a blue-shift. 
The estimated hydrogen column density of the low-energy absorber
ranges from $N_{\rm H}$=1.8 $\times$ 10$^{21}$~cm$^{-2}$
assuming the neutral absorption model (fit 2 of Table 1) 
to $N_{\rm H}$=1.1 $\times$ 10$^{22}$~cm$^{-2}$ assuming the ionized 
absorber model (fit 5 of Table 1); we note that these $N_{\rm H}$
values may have significant systematic errors due to the
inadequacy of the absorption models. Specifically, these models do not 
include effects such as possible velocity dispersion of the absorber,
BALs, and multiple ionization zones.
To obtain an estimate of the hydrogen column densities of the 
high-energy absorbers we performed a simple curve-of-growth analysis (Spitzer 1978) on the
detected absorption features. For no ionization correction (assuming all of the Fe is 
in the Fe ~{\sc xxv} ionization state)
and assuming solar abundances, the
total hydrogen column densities corresponding to the absorption lines  at 
$7.38_{-0.10}^{+0.17}$~keV
and $9.50_{-1.40}^{+0.90}$~keV 
are $N_{\rm H}$ = 4 $\times$ 10$^{22}$~cm$^{-2}$ and 
$N_{\rm H}$ = 4 $\times$ 10$^{23}$~cm$^{-2}$, respectively. 
We consider these estimates of the column densities 
also to have likely systematic errors
since the absorption lines may contain
multiple unresolved absorption components leading to an overestimate 
of the column densities of individual velocity components and
the $b$ parameters obtained from fits
of Gaussian lines to observed absorption features
are valid only in the optically thin limit.
On the other hand, possible saturation of the absorption lines will lead 
to an underestimate of the column densities.
Our preliminary comparison of the kinematic properties
of the low-energy and high-energy absorbers cannot
conclusively show whether they
are outflowing at similar velocities, however, our spectral 
analysis indicates that these
absorbers have different ionization parameters.
In particular, the best-fit ionization parameter for the low-energy absorption is
$\xi$ = $101_{-48}^{+58}$~erg~cm~s$^{-1}$ (see fit 5 of Table 1), whereas, the presence of highly
ionized Fe (high-energy absorption features) requires $\xi$ $\simgt$ 1000 erg~cm~s$^{-1}$.
We therefore conclude that the low and high-energy absorbers of \pgone\ are likely distinct;
however, higher spectral resolution is needed to confirm this result.

\subsection{Variability of X-ray BALs in \pgone\ and \apm}
We searched for variability of the X-ray BALs of \pgone\ and \apm\
by comparing their properties at different epochs.
We first compared the X-ray spectra of \pgone\ obtained 
in \chandra\ and \xmm\ observations separated by $\sim$ 19~weeks.
We fit the \chandra\ spectrum obtained
on 2000 June 2 for 26.8~ks simultaneously with the 
\xmm\ spectrum of \pgone\ using the spectral model of fit 5 of Table~1.
To account for the recently observed quantum efficiency decay of ACIS, 
possibly caused by molecular contamination of the ACIS filters, 
we have applied a time-dependent correction to the
ACIS quantum efficiency implemented in the \verb+XSPEC+ model \verb+ACISABS1.1+.
\footnote{ACISABS is an XSPEC model contributed to the \chandra\
users software exchange world wide web-site 
http://asc.harvard.edu/cgi-gen/cont-soft/soft-list.cgi} 
We allowed the normalization of the power-law 
component to be a free parameter in the fit
to account for any flux variability and/or systematic calibration uncertainties between
the \chandra\ and \xmm\ instrumental efficiencies.
We found a degradation in fit quality at the 90\%
confidence level compared to the fit of the same model to the \xmm\
spectrum of \pgone\ based on the $F$-test.
This change in fit quality is suggestive of variability of the
spectrum of \pgone. 

When we repeated the fit to the \chandra\ spectrum of \pgone\
allowing the model parameters of the low and high-energy 
absorbers to vary,
we obtain best-fit rest-frame line energies, widths, equivalent widths, 
and 68\% confidence levels   
of the absorption features of $E_{abs1} = 7.37_{-0.30}^{+0.14}$~keV,  
$\sigma_{abs1}$ $ < $ 0.35~keV, 
$EW_{abs1} = 96_{-73}^{+48}$~eV and $E_{abs2} = 8.5_{-0.2}^{+0.2}$~keV,
$\sigma_{abs2}$ $ < $ 0.25~keV, $EW_{abs2} = 178_{-92}^{+121}$~eV. 
The best-fit values for the ionization parameter and hydrogen column density of
the  low-energy ionized absorber during the \chandra\
observation are $\xi = 398_{-210}^{+260}$  erg~cm~s$^{-1}$ and 
$N_{\rm H} = 3.2_{-1.5}^{+2.1}$ $\times$ 10$^{22}$~cm$^{-2}$.
A comparison between the \chandra\ and \xmm\ observations indicates that
the line width of the second absorption feature varied to 
$\sigma_{abs2}  = 2.4_{-0.9}^{+1.2}$~keV
for the \xmm\ observation. The properties of the absorption line at $E_{abs1}$ = 7.4~keV
were consistent between both observations.
In Figures~4a and 4b we show
the $\Delta\chi$ residuals between the best-fit simple
absorbed power-law model and the \xmm\ PN 
and the \chandra\ ACIS spectra of \pgone, respectively.
The ionization parameter and hydrogen column density of
the  low-energy ionized absorber also appeared to vary
to $\xi = 101_{-48}^{+58}$  erg~cm~s$^{-1}$ and 
$N_{\rm H} = 1.1_{-0.4}^{+0.5}$ $\times$ 10$^{22}$~cm$^{-2}$ for the
\xmm\ observations.
In Figures~5a and 5b we show the 68\% and 90\% confidence contours of photon index
versus intrinsic low energy absorption column density 
and ionization parameter versus intrinsic column density 
for the \xmm\ and \chandra\ 
observations assuming fit 5 of Table 1.
These contours indicate a possible change
(at the 68\% confidence level) of the intrinsic column density
and the ionization parameter between observations.
To illustrate the significance of variability in the high-energy
absorption features detected in \pgone\ 
with \xmm\ and \chandra\ we present in Figures 6a and 6b  
confidence contours of $E_{abs2}$ vs. $\sigma_{abs2}$
and $E_{abs2}$ vs. line-flux, respectively.
We note, however, that these confidence contours are only suggestive of 
variability in the width and line flux of the absorption line at $E_{abs2}$
since this line is only detected at the 68\% confidence level in
the \chandra\ spectrum of \pgone. 
Flux and possibly spectral variability in the 0.2 -- 1.0~keV band
have been detected in earlier observations of this object performed 
with the {\it Einstein} IPC and {\it ROSAT} PSPC
on 1979 Dec 10 and 1991 Nov 21, respectively
(see Figures 5 and 6 of Chartas 2000). 
Specifically, the 0.2--2~keV flux varied from 6.5 $\times$ 10$^{-13}$ erg~s$^{-1}$~cm$^{-2}$
to 0.4 $\times$ 10$^{-13}$ erg~s$^{-1}$~cm$^{-2}$
between the {\it Einstein} IPC and {\it ROSAT} PSPC observations.
The 0.2--2~keV and 0.2--10~keV fluxes of \pgone\ observed with the PN 
are 2.7 $\times$ 10$^{-13}$ erg~s$^{-1}$~cm$^{-2}$ and 
6.1 $\times$ 10$^{-13}$ erg~s$^{-1}$~cm$^{-2}$, respectively.

We also searched for variability of the X-ray absorption lines 
recently detected in \chandra\ and \xmm\  observations of  the BALQSO \apm.
The \chandra\ and \xmm\ spectra
were fit with model 5 of Table 1.  Both fits were acceptable with 
reduced $\chi^{2}$ = 1.04 for 102 dof in the \chandra\ case 
and reduced $\chi^{2}$ = 1.04 for 116 dof  in the \xmm\ case.
A comparison between the two observations
separated by $\sim$ 1.8~weeks (proper-time) indicates that
the values (quoted with 68\% confidence errors) of 
$E_{abs1} = 8.05_{-0.04}^{+0.05}$~keV ,
$\sigma_{abs1} < 0.1$~keV , 
$E_{abs2} = 9.8_{-0.1}^{+0.1}$~keV, and
$\sigma_{abs2} = 0.44_{-0.1}^{+0.1}$~keV
observed during the \chandra\ observations of \apm\ varied to
$E_{abs1} = 8.15_{-0.14}^{+0.18}$~keV,
$\sigma_{abs1} = 0.6_{-0.2}^{+0.2}$~keV,
$E_{abs2} = 11.0_{-0.7}^{+0.7}$~keV, and
$\sigma_{abs2}  = 2.0_{-0.4}^{+0.3}$~keV 
for the \xmm\ observations. 
In Figures~7a and 7b we show
the $\Delta\chi$ residuals between the best-fit simple
absorbed power-law model and the \xmm\ PN 
and \chandra\ ACIS spectra, respectively.
The detection of significant variability of the widths and energies
of the high-energy absorption lines in \apm\
confirms the intrinsic origin of the high-energy absorption
features in this BALQSO (e.g., Hamann et al. 1995).
The fact that the two absorption lines 
were narrower during the \chandra\ observation than during the \xmm\ 
observation of \apm\ may explain
why these features were more easily detectable in the former case. 
We conclude that the observed variability of the energies and 
widths of the X-ray BALs in \apm\  over a timescale of 1.8~weeks (proper-time)
supports our earlier conclusion that these absorbers are most likely  
launched at relatively small radii of less than $10^{16}(M_{bh}/M_{8})^{1/2}$~cm.

To summarize, we have reported the detection of a relativistic
outflow in the BALQSO \pgone\ imprinted as resonant X-ray absorption lines
at rest-frame energies of  $7.38_{-0.10}^{+0.17}$~keV (90\% confidence)
and $9.50_{-1.40}^{+0.90}$~keV (68\% confidence).
These energies suggest the presence of two distinct absorbers outflowing at velocities
of $\sim$ 0.10$c$ and $\sim$ $0.34c$. We found significant variability of the properties of the X-ray
BALs in \apm\ over a period of  1.8~weeks (proper-time) 
supporting our earlier conclusion that X-ray BALs are launched at
relatively small radii compared to the expected UV BAL minimum
launching radii.

Resolving the low and high-energy absorption features seen from
\pgone\ and \apm\ is required to constrain better the 
kinematic, ionization, and absorbing
properties of the observed outflows. 
The X-ray calorimeter onboard the ASTRO-E2
satellite (scheduled to launch in early 2005)
has an effective area that is $\approx$ 4 times
larger than that of the \chandra\ gratings in
the energy range of the observed X-ray BALs
and an energy resolution of about 12~eV.
Future observations of \pgone\ and \apm\ with ASTRO-E2
may provide high-resolution spectra 
of X-ray BALs with exposure times of $\approx$ 300~ks.
The current plan for the $\it Constellation-X$ observatory
calls for a combination of four satellites that will 
provide the higher energy resolution and
larger effective area that is needed to 
resolve the complex structure of the X-ray BALs in \pgone\ and \apm\ 
with efficient exposure times of about 10~ks. 
In addition, monitoring of the X-ray BALs with $\it Constellation-X$
on short timescales of the order of days may allow us to track the acceleration
phase of the absorbers via the velocity shifts of the resonant absorption lines.

We thank those involved with the \xmm\ Guest Observer
Facility for assisting in the initial data reduction of \pgone. 
We acknowledge financial support from NASA grant NAG5-9949. 
WNB acknowledges financial support from NASA LTSA grant NAG5-13035 and
NASA grant NAG5-9932.

\clearpage

\normalsize

\beginrefer

\refer Arav, N., Li, Z., \& Begelman, M.~C.\ 1994, \apj, 432, 62 \\

\refer Arav, N., Barlow, T.~A., 
Laor, A., Sargent, W.~L.~W., \& Blandford, R.~D.\ 1998, \mnras, 297, 990 \\

\refer Arnaud, K.~A.\ 1996, ASP 
Conf.~Ser.~101: Astronomical Data Analysis Software and Systems V, 5, 17 \\

\refer Barlow, T. A., Hamann, F., \& Sargent, W. L. W. 1977,
in ASP Conf. Ser. 28, Mass Ejection from AGN, ed. N. Arav, I. Shlosman, \& R. J. Weymann,
(San Francisco:ASP), 13 \\

\refer Chartas, G.\ 2000, \apj, 531, 81 \\

\refer Chartas, G., Brandt, W.~N., Gallagher, S.~C., \& Garmire, G.~P.\ 2002, \apj, 579, 169 \\

\refer Chartas, G., Dai, X., \&  Garmire, G.~P. 2003, Carnegie Observatories 
Astrophysics Series, Vol. 2: Measuring and Modeling the Universe, 
ed. W. L. Freedman (Pasadena: Carnegie Observatories,\\ 
http://www.ociw.edu/ociw/symposia/series\\
/symposium2/proceedings.html) \\

\refer Done, C., Mulchaey, J.~S., Mushotzky, R.~F., \& Arnaud, K.~A.\ 1992, \apj, 395, 275 \\

\refer Gallagher, S.~C., Brandt, W.~N., Laor, A., Elvis, M., Mathur, S., Wills, B.~J., \& Iyomoto, 
N.\ 2001, \apj, 546, 795 \\

\refer Gallagher, S.~C., Brandt, W.~N., Chartas, G., \& Garmire, G.~P.\ 2002, \apj, 567, 37 \\

\refer Gallagher, S.~C., Brandt, W.~N., Chartas, G., Garmire, G.~P., \& Sambruna, R.~M.,
2003, Adv. Space Res., in press, astro-ph/0212304 \\

\refer Goodrich, R.~W.\ 1997, \apj,  474, 606 \\

\refer Green, P.~J., Aldcroft, T.~L., Mathur, S., Wilkes, B.~J., \& Elvis, M.\ 2001, \apj, 558, 109 \\

\refer Green, R.~F., Pier, J.~R., Schmidt, M., Estabrook, F.~B., Lane, A.~L., \& Wahlquist, H.~D.\ 
1980, \apj, 239, 483 \\

\refer Hamann, F., Barlow, 
T.~A., Beaver, E.~A., Burbidge, E.~M., Cohen, R.~D., Junkkarinen, V., \& 
Lyons, R.\ 1995, \apj, 443, 606 \\

\refer Hasinger, G., Schartel, N., \& Komossa, S.\ 2002, \apjl, 573, L77 (H02) \\ 

\refer Hewett, P. C. \& Foltz, C. B.\ 2003, in press \aj, astro-ph/0301191\\

\refer Impey, C.~D., Falco, 
E.~E., Kochanek, C.~S., Leh{\' a}r, J., McLeod, B.~A., Rix, H.-W., Peng, 
C.~Y., \& Keeton, C.~R.\ 1998, \apj, 509, 551 \\

\refer Krolik, J.~H.~\& Voit, G.~M.\ 1998, \apjl, 497, L5 \\

\refer Michalitsianos, A.~G., Oliversen, R.~J., \& Nichols, J.\ 1996, \apj, 461, 593 \\

\refer Proga, D., Stone, J.~M., \& Kallman, T.~R.\ 2000, \apj, 543, 686 \\

\refer Spitzer, L. 1978, Physical Processes in the Interstellar Medium (New York: Wiley) \\

\refer Stark, A.~A., Gammie, C.~F., Wilson, R.~W., Bally, J., 
Linke, R.~A., Heiles, C., \& Hurwitz, M.\ 1992, \apjs, 79, 77 \\

\refer Tolea, A., Krolik, J.~H., \& Tsvetanov, Z.\ 2002, \apjl, 578, L31 \\

\refer Tripp, T. M., Green, R. F., \& Bechtold, J.  1990, \apjl, 364, L29 \\

\refer Turnshek, D. A., Foltz, C. B., Grillmair, C. J.,
\& Weymann, R. J., 1988, ApJ, 325, 651 \\

\refer Verner, D.~A.~\& Yakovlev, D.~G.\ 1995, \aaps, 109, 125 \\

\refer Weymann, R.~J., Morris, S.~L., Foltz, C.~B., \& Hewett, P.~C.\ 1991, \apj, 
373, 23 \\

\endrefer

\clearpage
\scriptsize
\begin{center}
\begin{tabular}{clcc}
\multicolumn{4}{c}{TABLE 1}\\
\multicolumn{4}{c}{RESULTS FROM FITS TO THE \xmm\ SPECTRUM OF \pgone} \\
 & & &  \\ \hline\hline
\multicolumn{1}{c} {Fit$^{a}$} &
\multicolumn{1}{c} {Model} &
\multicolumn{1}{c} {Parameter$^{b}$} &
\multicolumn{1}{c} {Value$^{c}$} \\
  &                                                            &                                \\
1 & PL, neutral absorption        &$\Gamma$ &  1.91$_{-0.03}^{+0.03}$                 \\
  &  at source.                   & $N_{\rm H}$& (0.27$_{-0.05}^{+0.05}$) $\times$ 10$^{22}$~cm$^{-2}$  \\
  &                               & $\chi^2/{\nu}$              & 322.7/209                             \\
  &                               & $P(\chi^2/{\nu})$$^{d}$           & 7.1~$\times$~10$^{-7}$          \\
  &                                                             &           &                   \\
  &                               &                             &                                          \\
${2}$  & PL, neutral absorption   & $\Gamma$     & 1.76$_{-0.04}^{+0.04}$                           \\
  &  at source, and two           &  $N_{\rm H}$ & (0.18$_{-0.05}^{+0.05}$) $\times$ 10$^{22}$~cm$^{-2}$   \\
  &  Gaussian absorption lines    &  E$_{abs1}$       & 7.42$_{-0.2}^{+0.5}$~keV     \\
  &  at source.                    & $\sigma_{abs1}$   & $ < $ 0.55~keV ~$(68\%$~errors)      \\
  &                               &  EW$_{abs1}$      & 0.15$_{-0.11}^{+0.33}$~keV                      \\
  &                               &  E$_{abs2}$       & 9.9$_{-0.9}^{+0.7}$~keV               \\
  &                               &  $\sigma_{abs2}$  & 3.0$_{-0.5}^{+0.6}$~keV~$(68\%$~errors)                \\
  &                               &  EW$_{abs2}$      & 2.2$_{-0.8}^{+1.0}$~keV                    \\
  &                               &  $\chi^2/{\nu}$   & 267.4/201                                \\
  &                               &$P(\chi^2/{\nu})$$^{d}$ & 1.2~$\times$~10$^{-3}$                                 \\
  &                               &                   &                                          \\
${3}$   & PL, ionized absorption &$\Gamma$    & 1.94$_{-0.03}^{+0.05}$            \\
  & at source and absorption      & $N_{\rm H}$ & (1.2$_{-0.4}^{+0.5}$)  $\times$ 10$^{22}$~cm$^{-2}$          \\
  & edge.                         & $\xi$    &  $103_{-45}^{+58}$~erg~cm~s$^{-1}$                 \\
  &                               & $E_{edge}$    &  $7.07_{-0.23}^{+0.20}$~keV                  \\
  &                               & $\tau_{edge}$    &  $0.31_{-0.11}^{+0.12}$                  \\
  &                               & $\chi^2/{\nu}$    &   243.2/206                                       \\
  &                               &$P(\chi^2/{\nu})$$^{d}$ &   0.04                        \\
  &                               &                   &                                          \\
  &                               &                   &                                          \\
${4}$   & PL, ionized absorption &$\Gamma$    & 1.88$_{-0.02}^{+0.05}$            \\
  & at source and two absorption  & $N_{\rm H}$ & (1.0$_{-0.4}^{+0.5}$) $\times$ 10$^{22}$~cm$^{-2}$          \\
  & edges.                        & $\xi$    &  $91_{-46}^{+60}$~erg~cm~s$^{-1}$                 \\
  &                               & $E_{edge1}$    &  $7.03_{-0.32}^{+0.19}$~keV                  \\
  &                               & $\tau_{edge1}$    &  $0.31_{-0.11}^{+0.11}$                  \\
  &                               & $E_{edge2}$    &  $9.63_{-0.43}^{+0.77}$~keV                  \\
  &                               & $\tau_{edge2}$    &  $0.22_{-0.14}^{+0.14}$                  \\
  &                               & $\chi^2/{\nu}$    &   236.1/204                                       \\
  &                               &$P(\chi^2/{\nu})$$^{d}$ &   0.06                        \\
  &                               &                   &                                          \\
  &                               &                   &                                          \\
${5}$  & PL, ionized absorption   & $\Gamma$     & 1.90$_{-0.04}^{+0.04}$                           \\
  &  at source, and two           &  $N_{\rm H}$ & (1.1$_{-0.4}^{+0.5}$) $\times$ 10$^{22}$~cm$^{-2}$   \\
  &  Gaussian absorption lines    &  $\xi$           & 101$_{-48}^{+58}$~erg~cm~s$^{-1}$     \\
  &  at source.                   &  E$_{abs1}$       & 7.38$_{-0.10}^{+0.17}$~keV     \\
  &                               & $\sigma_{abs1}$   & $ < $ 0.44~keV     \\
  &                               &  EW$_{abs1}$      & $0.14_{-0.06}^{+0.06}$~keV                      \\
  &                               &  E$_{abs2}$       & 9.50$_{-1.4}^{+0.9}$~keV~$(68\%$~errors)              \\
  &                               &  $\sigma_{abs2}$  & 2.4$_{-0.9}^{+1.2}$~keV~$(68\%$~errors)       \\
  &                               &  EW$_{abs2}$      & $1.4_{-0.4}^{+0.6}$~keV                    \\
  &                               &  $\chi^2/{\nu}$   & 230.9/200                                \\
  &                               &$P(\chi^2/{\nu})$$^{d}$ & 0.07                                \\
  &                               &                   &                                          \\
\hline \hline
\end{tabular}
\end{center}
${}^{a}$All model fits include Galactic absorption toward the source (Stark et al. 1992).
All fits are performed on the combined spectrum of images A1, A2, B and C of PG~1115+080. \\
${}^{b}$All absorption-line parameters are calculated for the rest frame.\\
${}^{c}$All errors are for 90\% confidence unless mentioned otherwise with all
parameters taken to be of interest except absolute normalization.\\
${}^{d}$$P(\chi^2/{\nu})$ is the probability of exceeding $\chi^{2}$ for ${\nu}$ degrees of freedom
if the model is correct.\\

\clearpage
\begin{figure*}
\centerline{\includegraphics[width=14cm]{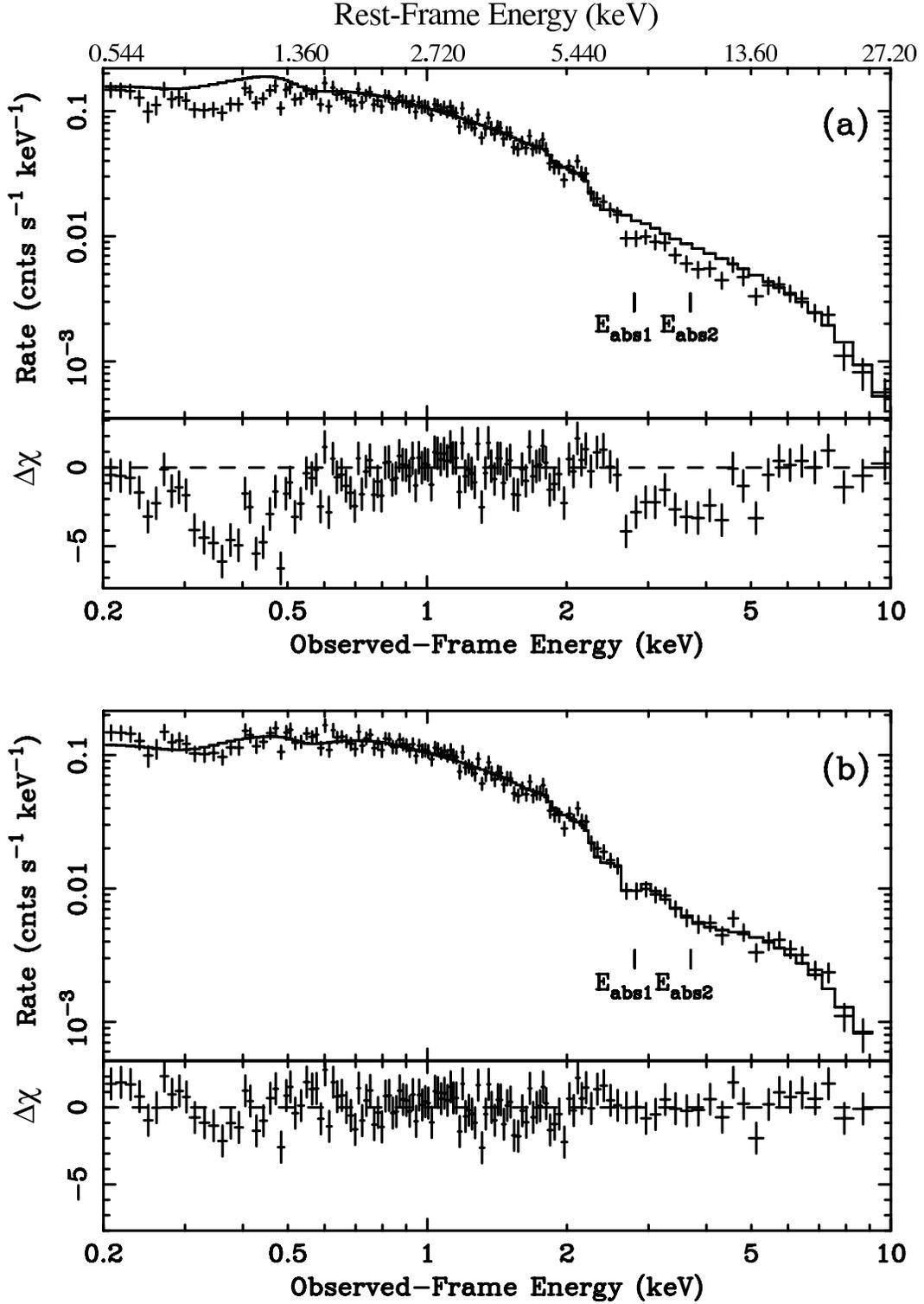}}
\caption{ \small  (a) The top panel shows the \xmm\ PN spectrum of
the combined images of \pgone\ fit with Galactic absorption and a 
power-law model to events with energies lying within 
the ranges of 1--2.6~keV and 5.2--10~keV. 
The lower panel shows the residuals of the fit in units of 1$\sigma$ deviations.
Two absorption features from 2.6--5.2~keV
are noticeable in the residuals plot.
(b) The top panel shows the PN spectrum
fit with Galactic absorption,
ionized absorption at the source,
a power-law model, and two Gaussian absorption lines. 
In the lower panel the residuals
indicate that this model can account for  
most of the spectral features of \pgone.
\label{fig1.eps}}
\end{figure*}

\clearpage
\begin{figure*}
\centerline{\includegraphics[width=14cm]{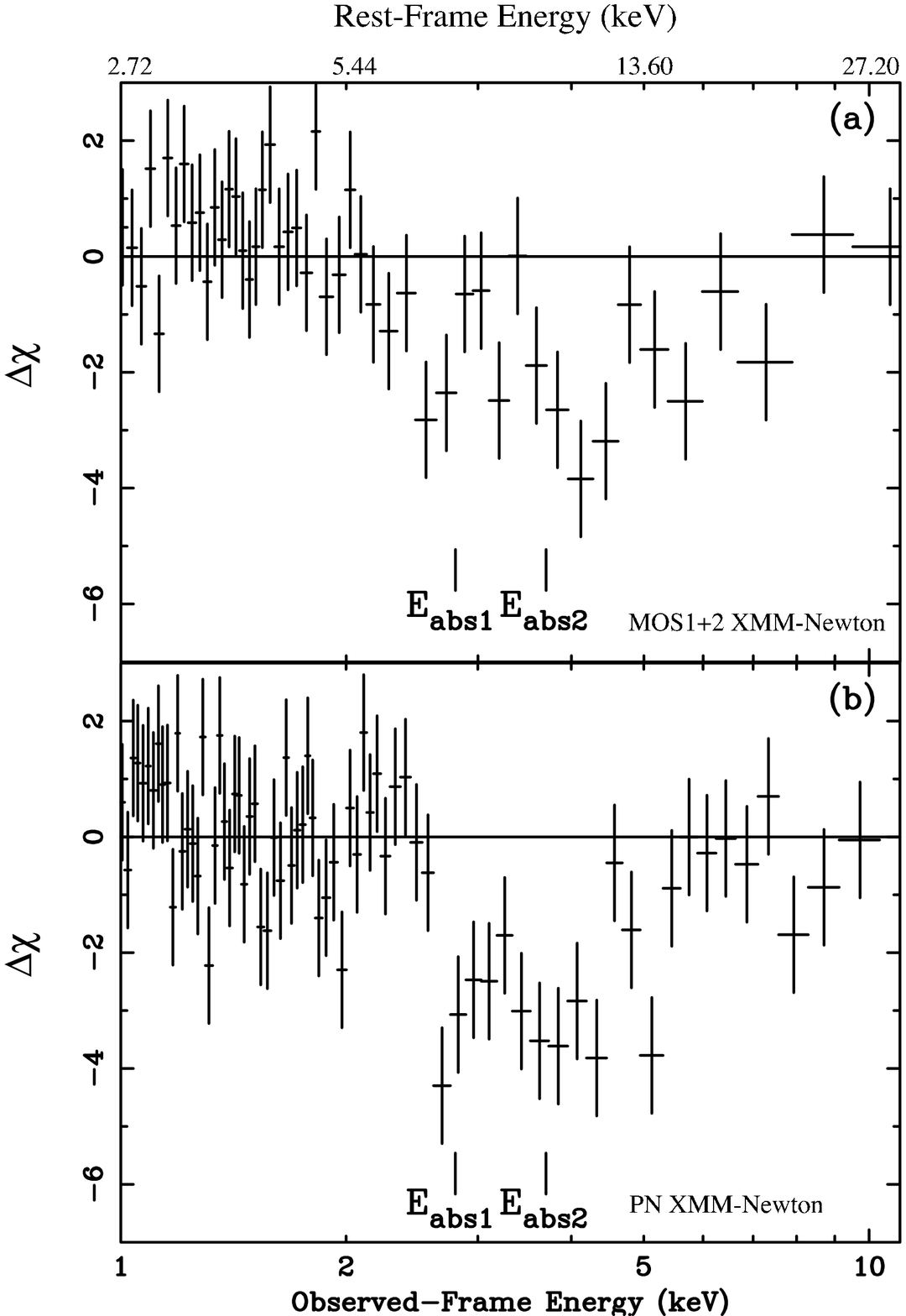}}
\caption{ \small  $\Delta\chi$ residuals between the best-fit 
Galactic absorption and power-law model and (a) the MOS1+2 spectrum of \pgone\ and
(b) the PN spectrum \pgone. This model is fit to events with energies
lying within the ranges 1--2.6~keV and 5.2--10~keV.
$E_{abs1}$ and $E_{abs2}$
 indicate the best-fit energies of the Gaussian absorption
lines obtained in fit 5 of Table~1.
\label{fig2.eps}}
\end{figure*}

\clearpage
\begin{figure*}
\centerline{\includegraphics[width=16cm]{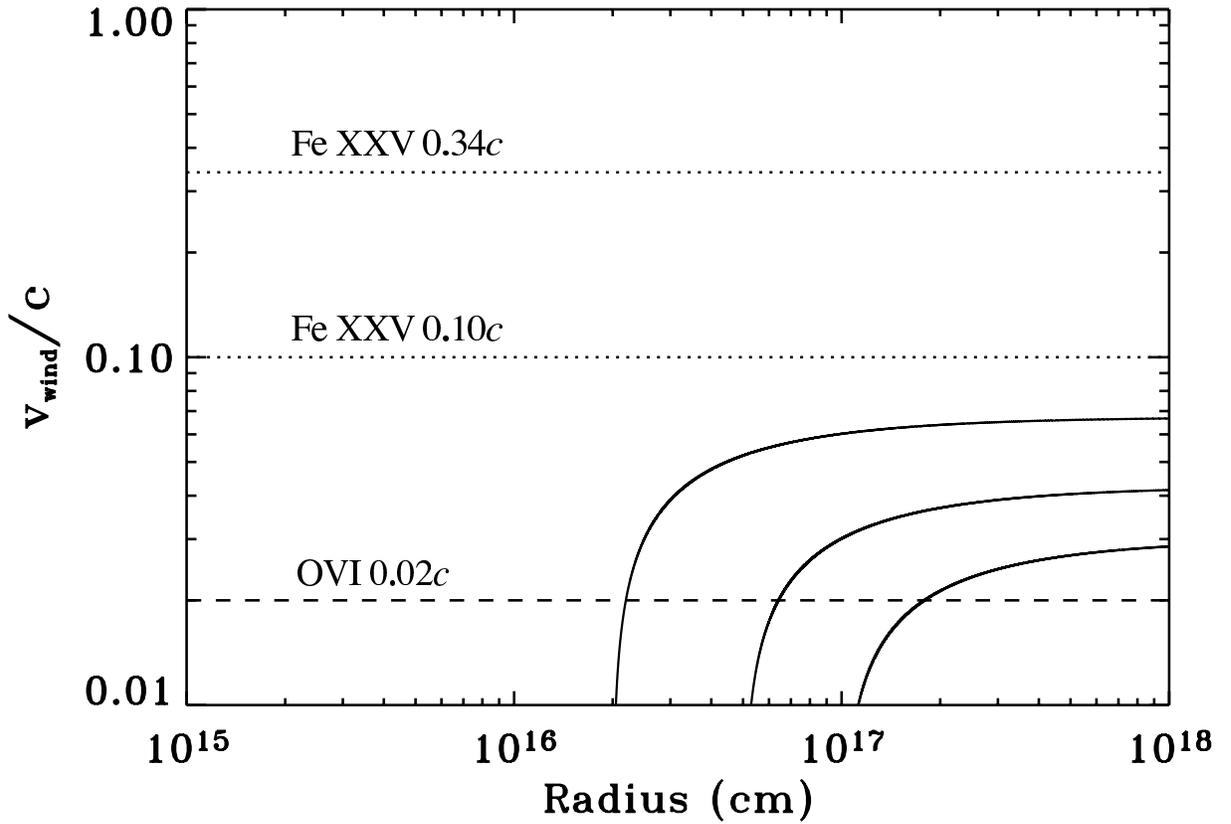}}
\caption{ \small Wind velocity as a function
of radius from the central source of \pgone\ for a radiation-pressure driven wind.
For a qualitative comparison we have estimated
the wind velocities  for launching 
radii of 2 $\times$ 10$^{16}$~cm, 5 $\times$ 10$^{16}$~cm,
and 1 $\times$ 10$^{17}$~cm. We have over-plotted
the observed O~{\sc vi} BAL (dashed line) and possible Fe~{\sc xxv} BAL (dotted lines) velocities.
\label{fig3.eps}}
\end{figure*}

\clearpage
\begin{figure*}
\centerline{\includegraphics[width=16cm]{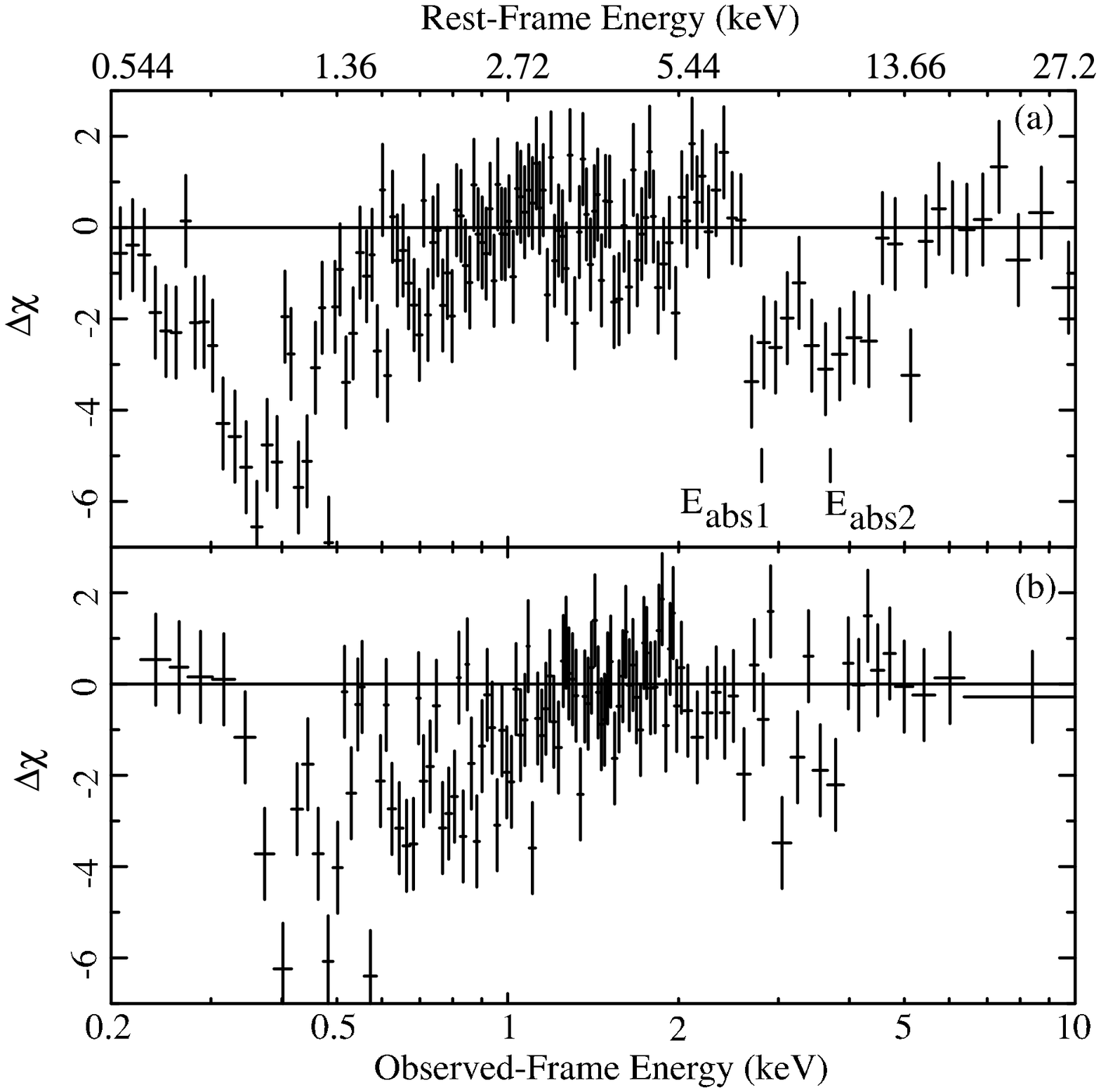}}
\caption{ \small $\Delta\chi$ residuals between the best-fit 
Galactic absorption and power-law model and (a) the \xmm\ PN spectrum of \pgone\ and
(b)  the \chandra\ ACIS spectrum of \pgone. 
The model has been fit to events with energies
lying within the ranges of 1--2.6~keV and 5.2--10~keV.
$E_{abs1}$ and $E_{abs2}$
indicate the best-fit energies of the Gaussian absorption
lines obtained using the spectral model described in fit 5 of Table~1.
\label{fig4.eps}}
\end{figure*}

\clearpage
\begin{figure*}
\centerline{\includegraphics[width=14cm]{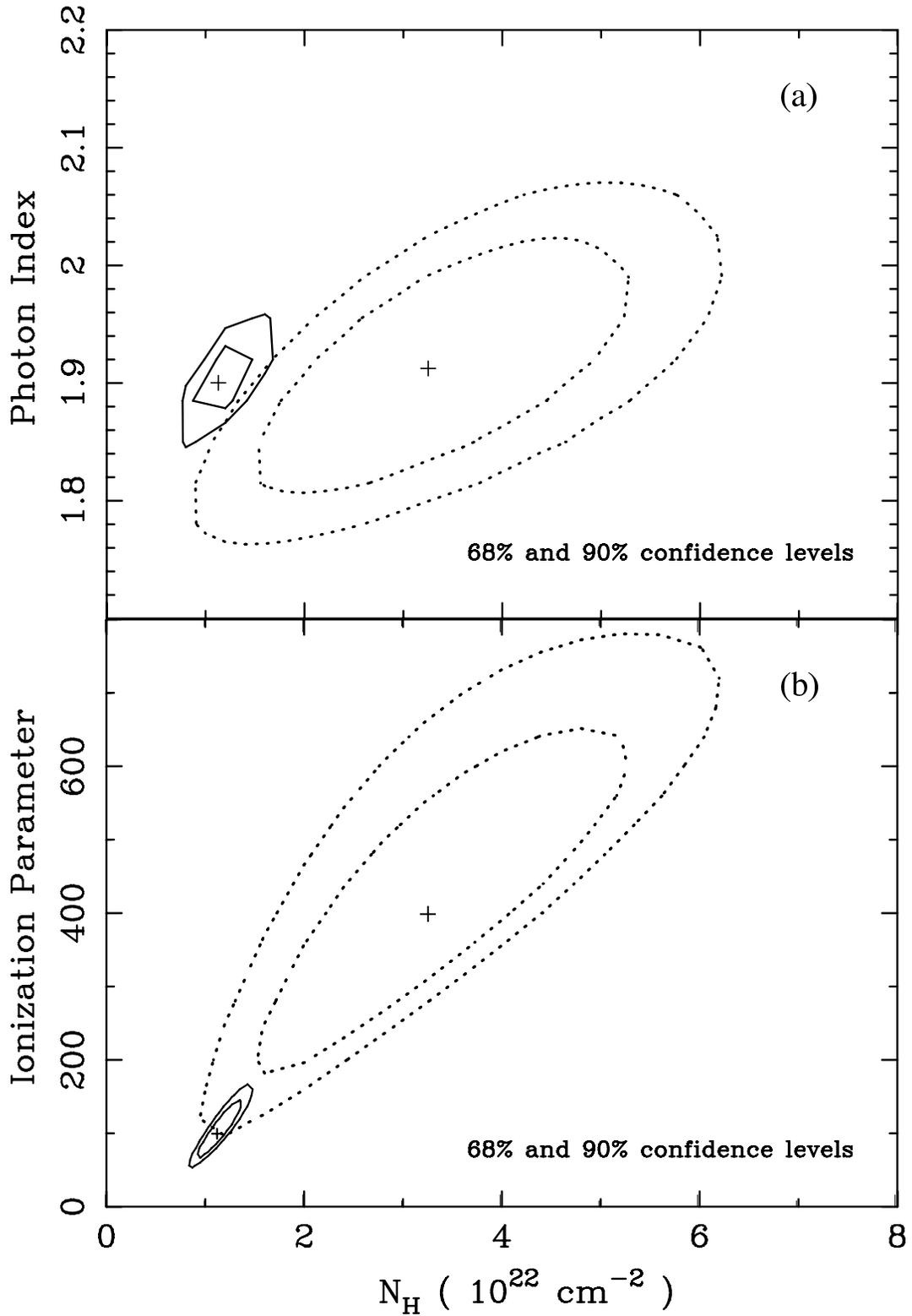}}
\caption{ \small 68\% and 90\%  confidence contours between 
(a) the spectral slope 
and the intrinsic absorption column density,
and (b) the ionization parameter and the intrinsic absorption column density
for the \xmm\ (solid contours) and \chandra\ 
(dotted contours) observations of \pgone\  assuming fit 5 of Table 1.
\label{fig5.eps}}
\end{figure*}

\clearpage
\begin{figure*}
\centerline{\includegraphics[width=14cm]{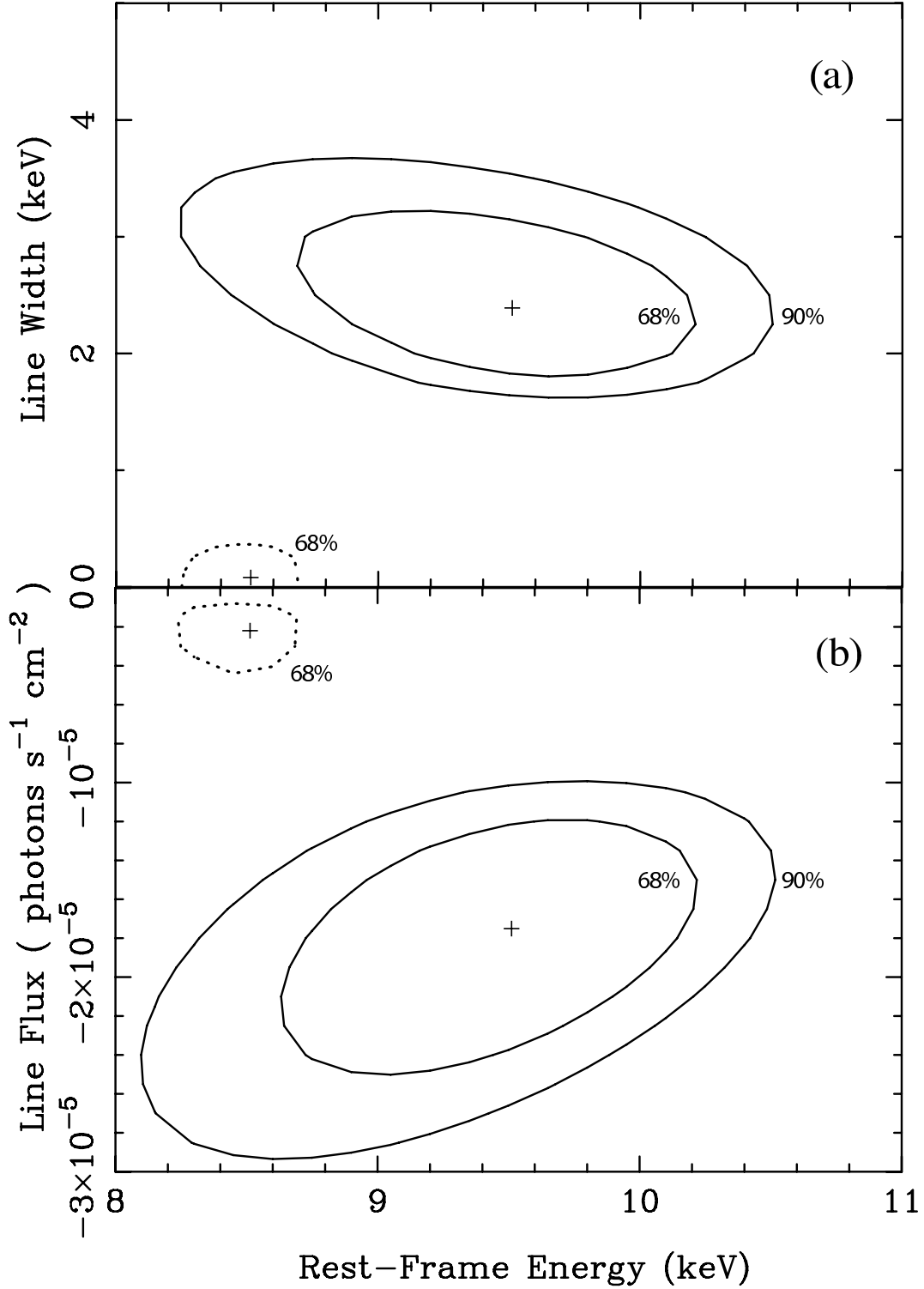}}
\caption{ \small (a) 68\% and 90\%  confidence contours between the
width of the absorption feature at $E_{abs2}$ 
and the energy $E_{abs2}$ for the \xmm\ (solid contours) and \chandra\
(dotted contours) observations of \pgone\ assuming fit 5 of Table 1.
(b) 68\% and 90\%  confidence contours between the flux 
of the absorption feature at $E_{abs2}$
and the energy $E_{abs2}$ for the \xmm\ (solid contours) and \chandra\
(dotted contours) observations of \pgone\ assuming fit 5 of Table 1.
These confidence contours are only suggestive of variability 
of the absorption line $E_{abs2}$ between the \xmm\ and \chandra\
observations since this absorption line is 
only detected at the 68\% confidence level
in the  \chandra\ spectrum.
\label{fig6.eps}}
\end{figure*}

\clearpage
\begin{figure*}
\centerline{\includegraphics[width=16cm]{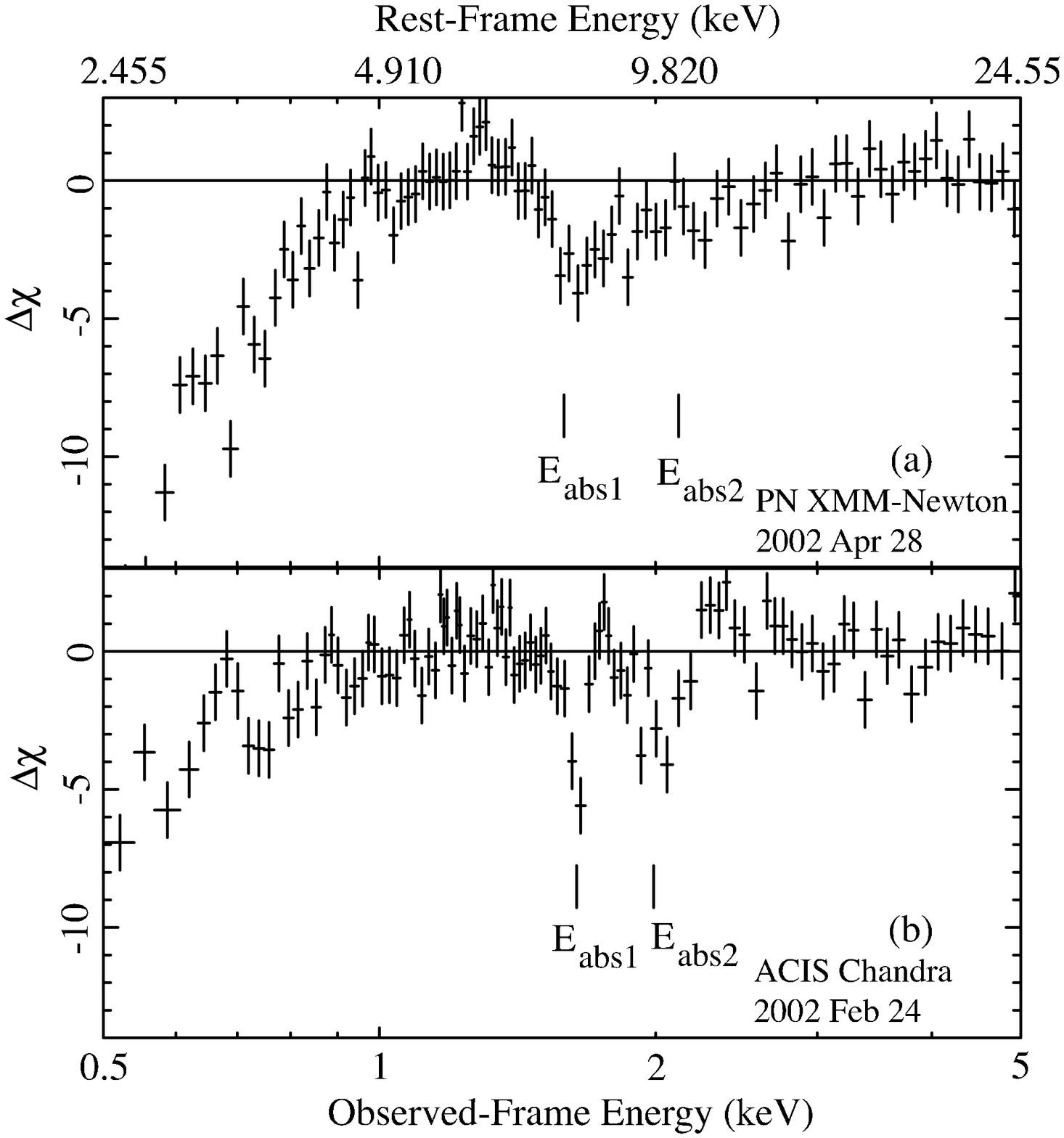}}
\caption{ \small $\Delta\chi$ residuals between the best-fit 
Galactic absorption and power-law model and (a) the \xmm\ PN spectrum and
(b)  the \chandra\ ACIS spectrum of \apm. 
This model is fit to events with energies
lying within the range 2.2--10~keV.
$E_{abs1}$ and $E_{abs2}$
indicate the best-fit energies of the Gaussian absorption
lines obtained using the spectral model described in fit 5
of Table~1.
\label{fig7.eps}}
\end{figure*}

\end{document}